# Physically-Based Inverse Rendering Framework for PET Image Reconstruction

Yixin Li, Soroush Shabani Sichani, Zipai Wang, Wanbin Tan, Baptiste Nicolet, Xiuyuan Wang, David A. Muller, Gloria C. Chiang, Wenzel Jakob and Amir H. Goldan

*Abstract*—Differentiable rendering has been widely adopted in computer graphics as a powerful approach to inverse problems, enabling efficient gradient-based optimization by differentiating the image formation process with respect to millions of scene parameters. Inspired by this paradigm, we propose a physically-based inverse rendering (IR) framework, the first ever platform for PET image reconstruction using Dr.Jit, for PET image reconstruction. Our method integrates Monte Carlo sampling with an analytical projector in the forward rendering process to accurately model photon transport and physical process in the PET system. The emission image is iteratively optimized using voxel-wise gradients obtained via automatic differentiation, eliminating the need for manually derived update equations. The proposed framework was evaluated using both phantom studies and clinical brain PET data acquired from a Siemens Biograph mCT scanner.

Implementing the Maximum Likelihood Expectation Maximization (MLEM) algorithm across both the CASToR toolkit and our IR framework, the IR reconstruction achieved a higher signal-to-noise ratio (SNR) and improved image quality compared to CASToR reconstructions. In clinical evaluation compared with the Siemens Biograph mCT platform, the IR reconstruction yielded higher hippocampal standardized uptake value ratios (SUVR) and gray-to-white matter ratios (GWR), indicating enhanced tissue contrast and the potential for more accurate tau localization and Braak staging in Alzheimer's disease assessment.

The proposed IR framework offers a physically interpretable and extensible platform for high-fidelity PET image reconstruction, demonstrating strong performance in both phantom and real-world scenarios.

*Index Terms*—MLEM, PET image reconstruction, physically-based inverse rendering

## I. INTRODUCTION

Positron Emission Tomography (PET) is a noninvasive imaging modality that enables quantitative measurement of biochemical and physiological processes by reconstructing the three-dimensional spatial distribution of radiotracers[1][2]. The early PET image reconstruction techniques were primarily based on analytical methods such as Filtered Backprojection (FBP) [3][4][5]. These methods assume an idealized imaging model and typically neglect various physical effects inherent to the PET acquisition process. As a result, analytical methods tend to amplify noise when it is applied in low-count scenarios and often yield suboptimal image quality [6][7][8]. The iterative reconstruction algorithms have been widely adopted due to their ability to model the imaging physics through the system matrix that defines the probabilistic mapping between image space and the detector pairs, which enable the improved spatial resolution and better contrast reconstructions [9][10][11].

Deep learning methods have shown significant promise in recent years for PET image reconstruction [12][13]. Broadly, these approaches can be categorized into two types. The first involves direct reconstruction, where a deep neural network is trained to map measured projection data (e.g., sinograms or list-mode events) directly to reconstructed images, effectively learning an implicit inverse mapping [14][15][16]. The second category encompasses deep learning–based iterative reconstruction, a hybrid strategy that integrates deep learning into conventional iterative algorithms [17][18][19]. In these methods, deep networks are typically embedded into the reconstruction loop to enhance image priors, perform denoising or refine intermediate solutions, while still preserving the physical and statistical models of PET.

This work was supported by the National Institute on Aging under grant U01AG082845 and by the National Institutes of Health under grant R01EB030413. (Yixin Li and Soroush Shabani Sichani are co-first authors) (Corresponding authors: Wenzel Jakob and Amir H. Goldan)

Yixin Li is with the Department of Radiology, Weill Cornell Medicine, New York, NY, 10065, USA. (yil4018@med.cornell.edu)

Soroush Shabani Sichani is with the Department of Radiology, Weill Cornell Medicine, New York, NY, 10065, USA, and with Department of Physics, Cornell University, Ithaca, New York, 14853, USA. (ss3325@cornell.edu)

Zipai Wang is with the Department of Radiology, Weill Cornell Medicine, New York, NY, 10065, USA. (ziw4004@med.cornell.edu)

Wanbin Tan is with the Department of Radiology, Weill Cornell Medicine, New York, NY, 10065, USA, and with Department of Biomedical Engineering, Stony Brook University, Stony Brook, NY, 11793, USA. (wat4006@med.cornell.edu)

Baptiste Nicolet is with School of Computer and Communication Sciences, École Polytechnique Fédérale de Lausanne, Lausanne, CH, 1015, Switzerland. (baptiste.nicolet@epfl.ch)

Xiuyuan Wang is with the Department of Radiology, Weill Cornell Medicine, New York, NY, 10065, USA. (xiw4004@med.cornell.edu)

David A. Muller is with School of Applied and Engineering Physics, Cornell University, Ithaca, NY, 14853, USA, and with Kavli Institute at Cornell for Nanoscale Science, Cornell University, Ithaca, NY, 14853, USA. (david.a.muller@cornell.edu)

Gloria C Chiang is with Department of Radiology, Brain Health Imaging Institute, Weill Cornell Medicine, New York, NY, 10065, USA. (gcc9004@med.cornell.edu)

Wenzel Jakob is with School of Computer and Communication Sciences, École Polytechnique Fédérale de Lausanne, Lausanne, CH, 1015, Switzerland. (wenzel.jakob@epfl.ch)

Amir H. Goldan is with the Department of Radiology, Weill Cornell Medicine, New York, NY, 10065, USA. (amg4017@med.cornell.edu)

However, the need for high-quality reference images paired with measured datasets for supervised training can be a significant challenge, particularly in clinical settings where such ground truth data are difficult or impossible to obtain. Moreover, generalization remains a critical concern because deep network trained on a specific dataset may fail when applied to data acquired under different conditions or from different patient populations. This domain shift can result in image artifacts or reconstruction errors that undermine the clinical reliability of the method [20][21].

Physically-based forward rendering methods incorporate the fundamental principles of light transport (e.g., reflection, refraction, and scattering) to simulate image formation processes with high physical fidelity [22][23]. Inverse rendering (IR) extends this concept to tackle a wide range of inverse problems in visual computing. It combines a differentiable image formation model with gradient-based optimization techniques to iteratively estimate the scene parameters (e.g., geometry, illumination, and material properties) to minimize discrepancies between simulated and observed data [24][25][26]. Recent advances in physically-based IR have introduced differentiable Monte Carlo simulations, enabling the computation of gradients with respect to complex and high-dimensional scene attributes, including heterogeneous volumetric structures and intricate surface properties [27][28][29]. This capability makes the IR especially powerful for solving problems where the analytical models are either intractable or nonexistent, thereby expanding its applicability to a wide range of challenging inverse imaging tasks.

Motivated by these advances, we extend the principles of IR to the domain of PET image reconstruction. In this context, the scene parameters correspond to the voxel-wise distribution of radiotracer activity within the subject, while the observations are the measured coincidence events recorded by the PET detector system. By formulating PET forward projection as a differentiable rendering process, we enable iterative optimization of the activity distribution to minimize the discrepancy between rendered and measured sinograms. This formulation forms the foundation of our proposed IR reconstruction framework, which tightly integrates physically accurate system modeling with automatic differentiation (AD). The framework supports end-to-end, gradient-based optimization without requiring explicit derivation of update equations, even for complex physics-driven models. Finally, we experimentally validate the proposed IR framework using both phantom studies and clinical human brain scan datasets to assess its practical effectiveness.

## II. RELATED WORKS

### A. PET Image Reconstruction

PET image reconstruction involves estimating the spatial distribution of a radiotracer's concentration within the field of view (FOV) of the scanner. The parameter vector $x \in \mathbb{R}^J$ contains the radiotracer concentration values across spatial points, where each voxel's image value is proportional to the total number of positron-emitting nuclei within the voxel volume, where $J$ denotes the number of voxels used to describe the reconstructed image [11].

The PET measurement process can be effectively modeled as a collection of independent Poisson random variables, capturing the noisy, discrete nature of photon detection. Given the discrete nature of the data and the approximately linear detection process, the relationship between the radioactive image and the expected values of true coincidence data detectable by each detector pair can be effectively modeled using a forward projection matrix

$$\bar{y} = Ax + r, \tag{1}$$

where $A \in \mathbb{R}^{I \times J}$ represents the system matrix, with $I$ and $J$ corresponding to the number of sinogram bins and PET image voxels, respectively. The element $a_{ij}$ denotes the probability of detecting an emission event from voxel $j$ by detector pair $i$. The additive term $r \in \mathbb{R}^I$ accounts for the random and scattered coincidence events. The mean number of the coincidence data $\bar{y} \in \mathbb{R}^I$ is related to the unknown tracer distribution through a transform that is defined by the physical characteristics of PET system [30].

The PET measured data, denoted as $y \in \mathbb{R}^I$, can be effectively modeled as a collection of independent Poisson random variables conditioned on the underlying tracer distribution $x$

$$p(y|x) = \prod_{i=1}^{I} \frac{\bar{y}_i^{y_i} e^{-\bar{y}_i}}{y_i!}. \tag{2}$$

The maximum likelihood estimate of the parameter vector $x$ can be computed by maximizing the log-likelihood of Poisson distribution

$$L(y|x) = \sum_{i=1}^{I}(y_i \log(\bar{y}_i) - \bar{y}_i - \log(y_i!)). \tag{3}$$

The EM algorithm is a general framework for solving this optimization problem [31]. The iterative update rule in Maximum Likelihood Expectation Maximization (MLEM) is given by

$$x^{k+1} = \frac{x^k}{A^T \mathbf{1}} A^T \left( \frac{y}{Ax^k + r} \right), \tag{4}$$

where $\mathbf{1} \in \mathbb{R}^I$ is a vector of ones, $k$ represents the iteration number, and $A^T$ is the transpose of the system matrix. This iterative process gradually improves the reconstructed image by incorporating corrections based on discrepancies between the measured and predicted data.

### B. Path-space Framework

Physically-based rendering algorithms simulate the behavior of light to generate realistic images from scene descriptions. Veach [32] introduced the path space formulation of light transport as a general framework for this task, treating the rendering problem as an integral over a high-dimensional space of light paths. This formulation decomposes the integration domain $\Omega$ into union of subspaces

$$\Omega = \bigcup_{n=2}^{\infty} \Omega_n. \tag{5}$$

A subset of paths involving exactly $n$ interactions can be expressed as:

$$\Omega_n = \{x_1 \cdots x_n | x_1, \cdots, x_n \in \mathcal{P}\}. \tag{6}$$

where $\mathcal{P}$ denotes the surfaces in the scene. The set $\Omega_n$ represents paths with $n$ successive interactions, describing the potential trajectories that light can take as they travel from the light source towards the scene, ultimately terminating at the sensor. The total illumination $I_j$ arriving at the virtual sensor can be expressed as a path integral of over the entire path space

$$I_j = \int_\Omega f(x)\,dA(x), \qquad (7)$$

where $f(x)$ is the contribution function quantifies the radiometric throughput along path $x$, and $dA(x)$ denotes the differential surface area measure across each vertex of the path.

III. METHODS

A. General Framework of Inverse Rendering for PET Image Reconstruction

IR is a powerful computational paradigm that bridges physically-based forward modeling with gradient-based optimization to solve complex inverse problems. The approach has been widely applied in the field of computer graphics (Fig. 1A), where the goal is to recover scene parameters (e.g., object geometry, material properties, and lighting conditions) by minimizing the discrepancy between rendered and observed images. This is accomplished through a differentiable rendering pipeline that simulates light transport and enables the computation of gradients with respect to scene parameters, which are then used within an optimization loop to iteratively refine the estimated parameters [27][33][34].

Inspired by this principle, we adapt the IR concept to PET image reconstruction (Fig. 1B). In this domain, the "scene parameters" correspond to the voxel-wise radiotracer activity distribution in the subject, and the forward rendering process simulates the detection of gamma-ray photons to generate sinograms. Our framework employs a hybrid differentiable forward model that combines Monte Carlo sampling with an analytical projector, capturing essential physical effects of the PET system such as attenuation, normalization and point spread function (PSF). The emission image is forward rendered into sinogram space, and the resulting synthetic sinogram is compared against the measured data using a differentiable objective function (e.g., Poisson log-likelihood or KL divergence).

The gradients of the loss function with respect to each voxel intensity are efficiently computed via AD, enabling quantification of how changes in voxel values impact the discrepancy between the rendered and observed sinograms. Finally, gradient-based optimization algorithms (e.g., Adam or L-BFGS) are employed to iteratively update the emission image by minimizing the loss function, thereby improving the fidelity of the reconstructed image.

All experiments in this study were conducted on a Linux workstation equipped with an NVIDIA RTX 6000 Ada GPU. Our physically-based IR framework was implemented on top of Dr.Jit, a high-performance just-in-time (JIT) compiler designed for differentiable rendering applications [35].

In the context of physically-based differentiable rendering, Monte Carlo simulations require the evaluation of millions of light or particle transport paths, each maintaining a substantial program state. Dr.Jit enables efficient execution of such simulations by compiling the entire Monte Carlo integration procedure into a single megakernel—a monolithic GPU kernel encapsulating all program instructions required to evaluate the integrand. This design minimizes memory traffic by storing most sample states in GPU registers and reduces overhead from inter-kernel communication, thereby improving computational throughput and scalability.

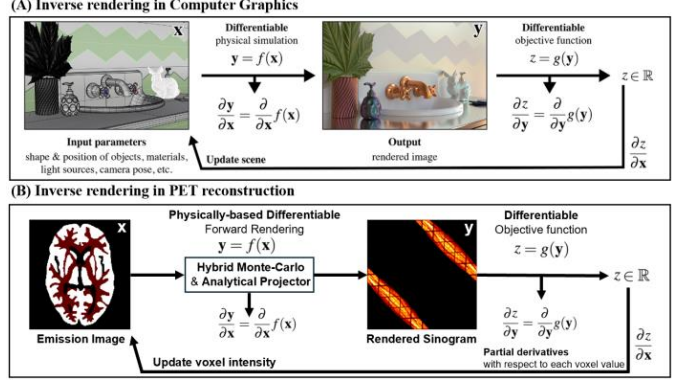

Fig. 1. Comparative illustration of IR in Computer Graphics and PET reconstruction. Fig. 1A is reproduced from [27].

B. Forward Rendering Modeling

The measured detector data in PET can be expressed as $m(c_i, c_j)$, representing the coincidence events detected along the line of response (LOR) formed between crystals $c_i$ and $c_j$. To simulate this process, Monte Carlo sampling is conducted within each scintillation crystal to sample points that represent the starting and ending points of multiple rays which are the possible gamma-ray paths through the radioactive tracer distribution. Sobol Monte Carlo sampling [36] is employed to uniformly sample points within the volume of each crystal. For each crystal pair $(c_i, c_j)$, the sampled points $p_i = (x_i, y_i, z_i)$ within $c_i$ and $p_j = (x_j, y_j, z_j)$ within $c_j$ define sub-LORs that collectively form the LOR between the crystals.

$$\begin{aligned}
x &\sim U\left(x_c - \frac{w}{2}, x_c + \frac{w}{2}\right), \\
y &\sim U\left(y_c - \frac{h}{2}, y_c + \frac{h}{2}\right), \\
z &\sim U\left(z_c - \frac{l}{2}, z_c + \frac{l}{2}\right),
\end{aligned} \qquad (8)$$

where $w, h, l$ are the width, height and length of the crystals.

To accurately model the intrinsic spatial resolution and PSF within the IR framework, we introduce a Gaussian perturbation kernel directly into the ray sampling process besides applying image-domain convolution. After incorporating the Gaussian perturbation $\delta \sim N(0, \sigma^2)$ into each coordinate independently, the distribution of the perturbed coordinates is given by the convolution of uniform and Gaussian distributions:

$$\begin{aligned}
x' &\sim U\left(x_c - \frac{w}{2}, x_c + \frac{w}{2}\right) * \mathcal{N}(0, \sigma^2), \\
y' &\sim U\left(y_c - \frac{h}{2}, y_c + \frac{h}{2}\right) * \mathcal{N}(0, \sigma^2), \\
z' &\sim U\left(z_c - \frac{l}{2}, z_c + \frac{l}{2}\right) * \mathcal{N}(0, \sigma^2).
\end{aligned} \qquad (9)$$

This perturbation effectively incorporates the system's spatial blurring directly into the forward projection model. The line passing through $p_i$ and $p_j$ can be expressed as

$$r_{ij}(t) = p_i + t(p_j - p_i), \qquad (10)$$

where $t \in [0, 1]$ parameterizes the line integral between the perturbed points.

$$\|r'_{ij}(t)\| = \sqrt{(x_j - x_i)^2 + (y_j - y_i)^2 + (z_j - z_i)^2}, \quad (11)$$

which is a magnitude of the derivative of $r_{ij}(t)$ to account for the physical distance along the line. Consequently, the sub-LOR along the perturbed line connecting $p_i$ and $p_j$ corresponds to a line integral $L_{ij}$ can be expressed as:

$$L_{ij} = \int_0^1 f(r_{ij}(t)) \|r'_{ij}(t)\| dt, \quad (12)$$

which captures the contribution of the radioactive tracer distribution to the detected coincidence events along that path.

When incorporating Time-of-Flight (TOF) information, the assumption of a uniform event distribution along the full LOR is replaced by a Gaussian probability distribution centered along the LOR. In practice, the TOF sinograms introduce an additional dimension to capture timing information, effectively partitioning the non-TOF sinogram into multiple TOF bins. Each bin collects coincidence events whose annihilation locations are most probable within a specific spatial range along the LOR, according to the corresponding TOF kernel. Each TOF kernel is modeled as a Gaussian distribution that describes the spatial probability of the annihilation events for zero measured time difference. The standard deviation of the Gaussian kernel corresponds to the timing resolution of the PET system [37][38]. To implement TOF information during the forward rendering, we introduce the Gaussian weighting function along each sub-LOR to represent the TOF measurement probability distribution

$$G_{TOF}(t, d_{TOF}, \sigma_{TOF}) = \frac{1}{\sqrt{2\pi\sigma_{TOF}^2}} \exp\left(-\frac{(d_{ij}(t) - d_{TOF})^2}{2\sigma_{TOF}^2}\right) \quad (13)$$

where $d_{ij}(t)$ is the scalar distance from the center of LOR to the point of annihilation, $d_{TOF}$ is the measured TOF kernel center spatial coordinate along the LOR, $\sigma_{TOF}$ represents the TOF resolution of the PET system. The modified sub-LOR integral with TOF Gaussian weighting is

$$L_{ij}^{TOF} = \int_0^1 f(r_{ij}(t)) G_{TOF}(t, d_{TOF}, \sigma_{TOF}) \|r'_{ij}(t)\| dt. \quad (14)$$

To compute the average line integral over all possible sub-LORs in continuous space, we normalize the total contribution by the product of the volumes of the crystal regions

$$\frac{1}{\int_{c_i} \int_{c_j} 1 dp_j dp_i} \int_{c_i} \int_{c_j} \int_0^1 f(r_{ij}(t)) G_{TOF}(t, d_{TOF}, \sigma_{TOF}) \|r'_{ij}(t)\| dt dp_j dp_i \quad (15)$$

For computational feasibility, this integral is approximated using a Monte Carlo estimator [28] during the forward rendering process. We generate a set of $B$ randomly sampled sub-LORs for each detector crystal pair, representing discrete photon interaction paths within the detector volumes. Each sub-LOR is evaluated using the integrand and their contributions are averaged to estimate the expected coincidence event count

$$L^{TOF} = \frac{1}{B} \sum_{b=1}^{B} L_{ij}^{TOF(b)}, \quad (16)$$

where $L_{ij}^{TOF(b)}$ denotes the line integral along the $b$-th sampled sub-LOR. This formulation leverages the probabilistic nature of gamma-ray interactions to model the distribution of coincidence events accurately.

The contribution function used in the forward rendering model is defined as

$$f(r_{ij}(t)) = W_e^i(r_{ij}(t)) \Phi(r_{ij}(t)) W_e^j(r_{ij}(t)), \quad (17)$$

where $\Phi(\cdot)$ denotes the radiotracer distribution within the activity image. $W_e^{i(j)}(\cdot)$ is the binary indicator function based on position that equals 1 when the interaction occurs inside the crystal, and 0 otherwise.

To ensure the rendered sinogram reflects the physical processes of PET imaging, we incorporate normalization and attenuation corrections. The normalization correction accounts for variations in system sensitivity, including crystal efficiency and geometric factors. It is estimated using a uniformly radioactive cylindrical phantom acquired under the same system configuration. The attenuation correction compensates for photon absorption along the LORs, which is implemented by multiplying the rendered sinogram with a precomputed attenuation correction factor derived from the attenuation map.

$$\mathcal{R}(\mathbf{I}) = L_R \cdot L_N \cdot L_{AC}, \quad (18)$$

where $\mathcal{R}(\mathbf{I}) \in \mathbb{R}^M$ denotes the corrected sinogram corresponding to the emission image $\mathbf{I} \in \mathbb{R}^N$. $L_R \in \mathbb{R}^M$ represent the raw rendered sinogram based on the hybrid Monte Carlo and analytical projector. $L_N, L_{AC} \in \mathbb{R}^M$ demonstrate the normalization and attenuation correction coefficient maps, respectively.

### C. Inverse Rendering Modeling

The inverse process in PET image reconstruction aims to refine the emission image $\mathbf{I}$ iteratively by minimizing the discrepancy between the rendered sinogram generated during forward rendering and the actual measured sinogram data. To achieve this, an objective function $\mathcal{L}(\mathbf{M}; \mathcal{R}(\mathbf{I}))$ is defined, where $\mathbf{M} \in \mathbb{R}^M$ represents the measured sinogram and $\mathcal{R}(\mathbf{I})$ is the rendered sinogram map from the forward rendering function. This function quantifies the alignment between the two sinograms and serves as a measure of reconstruction accuracy. The optimal reconstructed image $\hat{\mathbf{I}} \in \mathbb{R}^N$ is obtained by solving the following optimization problem

$$\hat{\mathbf{I}} = argmin_{\mathbf{I}} \mathcal{L}(\mathbf{M}; \mathcal{R}(\mathbf{I})), \quad (19)$$

where $\mathcal{L}(\cdot;\cdot)$ represents the loss function that evaluates the reconstruction accuracy of the rendering operator. To refine the emission image iteratively, the update form of the image estimate can be generically expressed as

$$\mathbf{I}^{k+1} = \mathbf{I}^k - \alpha^k \mathbf{J}^k, \quad (20)$$

where $\alpha^k$ represents a step size parameter at iteration $k$, determined either through a fixed schedule or adaptive strategy. $\mathbf{J}^k \in \mathbb{R}^N$ denotes a search direction derived from the local system response, typically chosen as the gradient of the loss function with respect to the image [9].

$$\mathbf{J}^k = \nabla_{\mathbf{I}} \mathcal{L}(\mathbf{M}; \mathcal{R}(\mathbf{I}^k)). \quad (21)$$

Utilizing the Dr.JIT backend, our platform supports AD that enables the computation of $\nabla_{\mathbf{I}} \mathcal{L}$ directly from the forward model definition, which eliminates the need for explicit gradient derivation and facilitates the inclusion of the complex physics components during the forward model $\mathcal{R}(\cdot)$.

Considering the maximum likelihood estimation in an alternative way is to use gradient-based optimization procedures, where MLEM algorithm can be rewritten as a gradient ascent expression [39].

$$\mathbf{I}^{k+1} = \mathbf{I}^k + \frac{\mathbf{I}^k}{\mathbf{I_s}} \cdot \frac{\partial \mathcal{L}}{\partial \mathbf{I}^k}, \quad (22)$$

where $\mathbf{I_s}$ represents the sensitivity image that compensates the non-uniform spatial detection efficiency among the PET systems. We define $\mathbf{I_s}$ as the gradient of the rendered sinogram with respect to a uniform input image $\mathbf{I_o}$

$$\mathbf{I_s} = \nabla_{\mathbf{I_o}} \left( \mathbf{1}^\mathrm{T} \mathcal{R}(\mathbf{I_o}) \right) = \left( \frac{\partial \mathcal{R}(\mathbf{I_o})}{\partial \mathbf{I_o}} \right)^\mathrm{T} \mathbf{1}, \quad (23)$$

where $\frac{\partial \mathcal{R}(\mathbf{I_o})}{\partial \mathbf{I_o}} \in \mathbb{R}^{M \times N}$ denotes the Jacobian matrix of the forward operator $\mathcal{R}$. This gradient-based definition reflects the voxel-wise contribution to the rendered sinogram and is directly compatible with invers rendering pipelines.

IV. EXPERIMENTS

A. Ultra-micro Derenzo Phantoms

The ultra-micro Derenzo phantom (Data Spectrum Corporation) containing hot-spot arrays with diameters of 0.75, 1.00, 1.35, 1.70, 2.00, and 2.40 mm was used for evaluating reconstructed images performance from different platforms. The phantom was filled with 37 MBq (1 mCi) of $^{18}$F-FDG and positioned at the center of the FOV of the Prism-PET scanner [40][41][42] with a total of 60 minutes acquisition time. The acquired list-mode data were post-processed to filter inter-crystal scatter (ICS) events, photoelectric correction using depth-of-interaction (DOI) information and normalizing for crystal efficiency variations. Image reconstruction was performed with approximately 30 million events using both the CASToR platform [43] and our IR framework. In both cases, the MLEM algorithm was applied with 200 iterations. A uniform voxel size of 0.5×0.5×0.5 mm³ was used during reconstruction. PSF modeling was incorporated using a 3D Gaussian kernel with a full width at half maximum (FWHM) of 1.2×1.2×1.2 mm³. Quantitative evaluation was performed by calculating the signal-to-noise ratio (SNR) within each hot-spot region, defined as the ratio between the mean voxel intensity and the standard deviation. Additionally, peak-to-valley ratio (PVR) was computed to further assess contrast and resolution performance.

B. 3D Hoffman Brain Phantom

The 3D Hoffman brain phantom was filled with 37 MBq (1 mCi) of $^{18}$F-FDG and positioned at the center of the FOV of Prism-PET. The phantom was scanned for 240 minutes. List-mode data were processed using a pipeline that included ICS rejection, DOI rebinning, time offset correction, normalization, and attenuation correction. Approximately 50 million counts coincidence events were recorded, and image reconstruction was conducted using both the CASToR platform and the IR framework, employing the MLEM algorithm with voxels of 0.5×0.5×0.5 mm³. Additionally, an isotropic 3D gaussian of 1.5 mm FWHM was applied during the reconstruction.

To assess image quality and quantification accuracy, two metrics were computed: ventricle spill-over ratio (SOR) and the coefficient of variation (COV) within the gray matter regions. The ventricle SOR defined as the mean intensity in the ventricle region normalized by the mean intensity in the surrounding gray matter, serves as an indicator of partial volume and resolution effects. A lower SOR indicates better separation between low- and high-uptake regions. The COV within the gray matter, defined as the standard deviation divided by the mean voxel intensity, quantifies noise or heterogeneity in presumably uniform uptake regions. A lower COV indicates higher image uniformity and less noise.

C. Human Subject Study

A human subject was enrolled under an approved Institutional Review Board (IRB) protocol from Weill Cornell Medicine (WCM) to investigate the feasibility and effectiveness of the IR framework. The human subject was administered approximately 5 mCi [18F] MK6240, The PET scan was performed 60 minutes post-injection.

PET images were reconstructed by Siemens Biograph mCT with 4 iterations, 21 subsets OSEM, and 2×2×2 mm³ voxels. The IR framework used the same voxel resolution and employed 84 iterations of the MLEM algorithm. The subject also underwent a 3D T1-weighted MRI scan on a 3T Siemens PRISMA scanner for co-registration and segmentation. The hippocampus was delineated as the region of interest (ROI) in the PET images, and standardized uptake value ratios (SUVR) were computed using the cerebellar gray matter as the reference region. COV within gray matter was used to evaluate image noise. The gray matter to white matter uptake ratio (GWR) was calculated to assess image contrast.

V. RESULTS

A. Ultra-micro Derenzo Phantoms

Fig. 2 presents the reconstructed images of the ultra-micro Derenzo phantom using both the CASToR toolkit and the proposed IR framework. All hot-spots with a diameter of 1.00 mm or larger can be clearly resolved across the entire FOV for the phantom in both reconstructions.

Fig. 3 demonstrates the quantitative comparison of SNR and PVR across different hot-spot diameters. As shown in the SNR plot, the IR reconstruction consistently yields higher SNR values compared to CASToR across all spot sizes, indicating improved noise suppression and stability in small volume quantification.

Conversely, the PVR analysis reveals that CASToR exhibits slightly superior spatial contrast recovery for diameters ≥1.35 mm with higher PVR values compared to IR. This suggests that CASToR better preserves edge definition and background delineation, likely due to its sharper but potentially noisier reconstruction profile. However, IR achieves a 12.33% higher PVR than CASToR for the 1 mm diameters hot spots demonstrating superior recovery of faint structures.

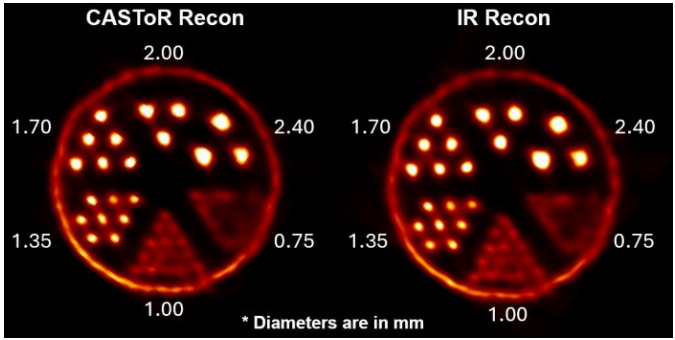

Fig. 2. Reconstructed images of ultra-micro Derenzo phantoms using the CASToR reconstruction toolkit and the proposed IR framework.

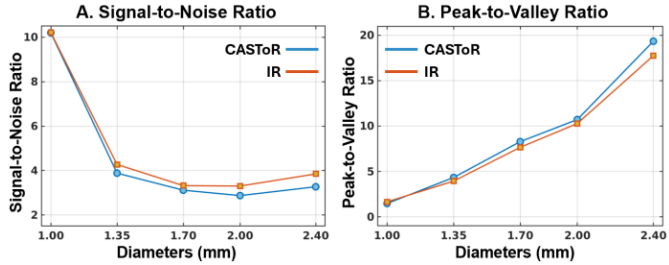

Fig. 3. Signal-to-Noise ratio and Peak-to-Valley ratio plotted across varying hot-spot diameters in the ultra-micro Derenzo phantom.

### B. 3D Hoffman Brain Phantom

Fig. 4 shows representative axial slices of the reconstructed 3D Hoffman brain phantom using the CASToR and IR platforms. The IR reconstruction demonstrates the improved anatomical contrast and more uniform intensity distributions within both gray and white matter regions. As shown in Table I, IR reconstruction achieves a lower COV of 7.84% compared to 8.01% in CASToR, demonstrating improved noise characteristics and signal stability across gray matter structures. IR reconstruction also achieves a lower SOR ($0.1167 \pm 0.0005$) compared to CASToR ($0.1243 \pm 0.0011$), suggesting improved suppression of activity spill-in from gray matter to ventricles and hence better spatial fidelity.

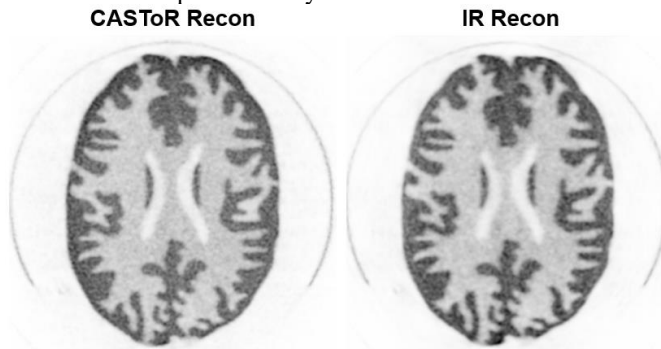

Fig. 4. Representative axial slices of the reconstructed Hoffman brain phantom obtained using the CASToR reconstruction toolkit and the proposed IR framework.

TABLE I
QUANTITATIVE ANALYSIS OF HOFFMAN BRAIN PHANTOM

|  | CASToR | IR |
|---|---|---|
| COV | 8.013% | 7.842% |
| SOR | $0.1243 \pm 0.0011$ | $0.1167 \pm 0.0005$ |

### C. Human Subject Study

A qualitative visual comparison of the Siemens Biograph mCT reconstruction and the proposed IR framework is presented in Fig. 5, where the IR image appears to show the enhanced spatial resolution and sharper delineation of cortical structures.

Quantitative analysis revealed that the IR reconstruction yielded higher regional uptake values and tissue contrast compared to the Siemens image (Table II). Specifically, the IR framework produced a 9.97% higher SUVR in the right hippocampus and a 3.52% higher GWR. However, IR reconstruction demonstrates a 3.83% higher COV in the gray matter, suggesting a slightly increased noise level despite the improved contrast.

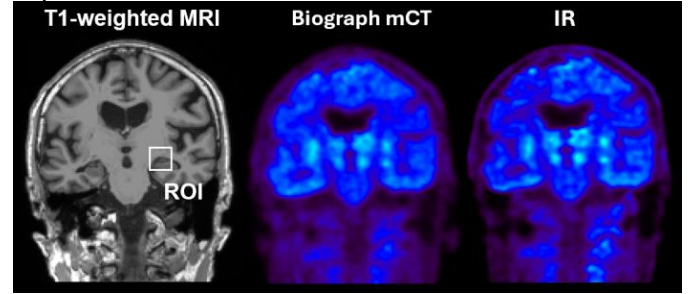

Fig. 5. Comparison of PET reconstruction methods in a human subject with corresponding T1-weighted MRI. The coronal views show anatomical reference, PET image from Siemens Biograph mCT reconstruction, and PET image generated by IR framework. The white square highlights the hippocampus that is designated as the target ROI for SUVR analysis.

TABLE II
QUANTITATIVE ANALYSIS OF HUMAN SUBJECT

|  | Biograph mCT | IR |
|---|---|---|
| COV | 13.23% | 17.06% |
| SUVR | $0.7932 \pm 0.1162$ | $0.8723 \pm 0.1311$ |
| GWR | $1.2537 \pm 0.2252$ | $1.2978 \pm 0.2776$ |

### D. Benchmarking Inverse Rendering

Here we compare the IR framework with the state-of-the-art CASToR. The IR reconstruction was performed using an Nvidia RTX 6000 Ada with 48Gb of memory. CASToR was run on the same workstation, with 2× Xeon E5-2699 v4 (44C/88T) with 2 NUMA nodes and L3 cache 56 MB per socket. Figure 6 illustrates the reconstruction time for the Hoffman and Hotspot phantoms reconstructed with IR and CASToR. We averaged 10 reconstruction times (sum of Projection and Optimization) for each phantom. Overall, IR completed one reconstruction of the Hoffman brain phantom in 1440.3s while it took CASToR 4.5x time to perform the same task (6373.0s). In the Hotspot phantom we observe almost 9.75x boost (243.7s compared to 2374.0s) in speed when we reconstruct using IR platform compared to CASToR. This almost twofold boost for the Hotspot phantom compared to Hoffman primarily reflects the image size variation. A summary of the reconstruction parameters and implementation details are included in table III.

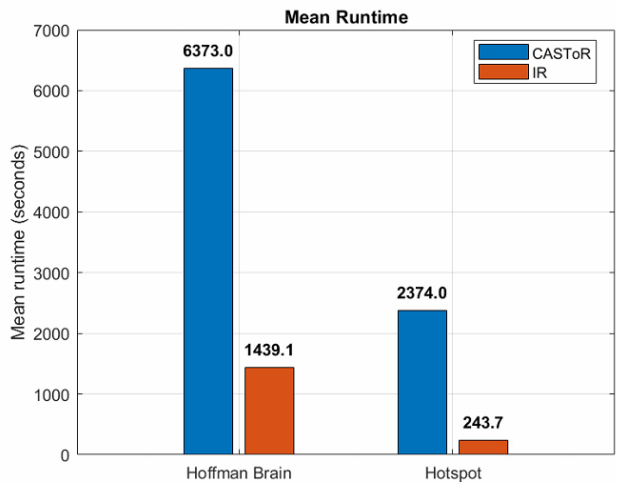

Fig. 6. Mean run time for Hoffman Brain and Hotspot phantoms. Inverse Rendering outperforms CASToR by 4.4x and 9.7x on Hoffman and Hotspot phantoms respectively.

TABLE III
Reconstruction Parameters used in CASToR and IR

|  | Hoffman | Hotspot |
| --- | --- | --- |
| PSF FWHM | 1.5 | 1.2 |
| Number of iterations | 80 | 200 |

## VI. Discussion

The IR reconstruction demonstrated superior noise suppression, as evidenced by a significantly lower COV in the gray matter of the Hoffman phantom and a higher SNR in the hot-spot phantom than CASToR reconstruction. Moreover, the lower ventricle SOR for IR indicates better recovery of cold regions, which suggests that IR would be more reliable for delineating non-avid region. However, this emphasis on noise suppression comes at a measurable cost to high-frequency spatial resolution. Compared with CASTOR reconstruction, the slightly lower PVR in IR can be attributed to the aggressive noise reduction strategy. This process reduces the peak amplitude of the hot spots and fills in the valleys between them, thereby suppressing noise at the expense of the fine-feature separation measured by PVR.

A comparison of the clinical PET images indicated that the IR framework yielded superior tissue contrast and regional quantification compared to the standard Siemens Biograph mCT platform. The IR reconstruction produced a higher hippocampal SUVR and an increased GWR, suggesting more effective mitigation of partial volume effects and enhanced preservation of fine anatomical details. However, a slightly higher COV was observed in the gray matter, indicating that the gain in contrast may came at the cost of slightly increased image noise. Nevertheless, this trade-off appears favorable for applications where structural clarity and regional quantification are prioritized over minimal image noise. Ultimately, the superior resolution and quantitative accuracy demonstrated by our IR framework may enable a more precise localization of tau pathology, potentially improving the reliability of in vivo Braak staging in the clinical evaluation of Alzheimer's disease.

Compared to conventional implementations of statistical reconstruction models, one of the most significant distinctions of IR lies in how the forward model is constructed and used during the iterative process. Traditional implementations rely on an explicitly defined system matrix that characterizes the detection probability of emissions from individual voxels being detected at corresponding sinogram elements. This matrix is typically derived from a combination of analytical and empirical models and remains fixed throughout the iterative reconstruction [44]. In contrast, the system model of IR framework is embedded directly into a differentiable and physics-driven rendering process. The forward projection is performed via path tracing, where the line integrals along the trajectories of coincident gamma rays are calculated dynamically using the actual geometry and physical properties of the scanner, which allows for a voxel-wise simulation of photon transport without requiring an explicit system matrix.

Moreover, the IR framework enhances physical realism by modeling the interaction of photons within the detector crystals using Monte Carlo simulations. By sampling random interaction points inside the scintillator volume, multiple sub-LORs are generated for each detector pair. Averaging over these sub-paths results in line integrals that better approximate the measured data, particularly accounting for effects such as depth-dependent blurring [45]. This dynamic and data-driven modeling capability enables the IR framework to more accurately represent the physics of PET detection compared to fixed matrix-based methods.

Our Dr.Jit implementation of Inverse Rendering shows a substantial performance improvement over CASToR. This is mainly because unlike the conventional PET reconstruction algorithms that incorporate parallelization using OpenMP for shared-memory CPU execution and MPI for distributed memory parallelization, reconstructions on GPU's allow for more parallelization and higher memory bandwidth. In addition, due to Dr.Jit's JIT-fused CUDA kernels, we can further minimize device-memory latency [46].

Another unique strength of the IR framework lies in its fully differentiable optimization pipeline. The voxel-wise activity distribution is optimized through a gradient-based objective function that quantifies the discrepancy between rendered and measured sinograms. This is achieved by using the AD to enable the efficient and accurate computation of partial derivatives of the loss function with respect to voxel values via backpropagation. Therefore, this framework eliminates the need for manually deriving update equations for different objective functions, such as least squares, weighted least squares, KL divergence, or Poisson log-likelihood. The AD mechanism closely parallels deep learning–based iterative PET reconstruction approaches, particularly self-supervised learning models [47]. However, unlike deep learning–based models that rely on architectural design choices and training data quantity for generalization, the IR framework offers a fully interpretable and physically grounded reconstruction pipeline. Reconstructions are inherently patient-specific and do not suffer from domain shift or generalization issues commonly observed in data-driven methods.

Furthermore, the inherent flexibility of the physics-based forward model in the IR framework allows for the seamless incorporation of increasingly complex physical effects into the reconstruction process. The factors such as Compton scattering, positron range blurring and annihilation photon acollinearity can be directly modeled through Monte Carlo simulations embedded in the forward rendering pipeline. Convolving these effects into the forward rendering process could improve the physical fidelity of the simulated data and enhance the reconstruction accuracy.

As the forward model becomes more comprehensive, the mathematical expression linking image parameters to the measured data becomes significantly nonlinear and high-dimensional, making manual derivation of gradients for optimization purposes impractical. However, the use of AD overcomes this bottleneck by enabling accurate and scalable gradient computation, regardless of the underlying model complexity. This capability ensures that the IR framework remains practical and efficient even when modeling the full physical behavior of the PET acquisition process. Moreover, this extensibility makes IR a powerful framework not only for improving image quality but also for exploring and potentially addressing fundamental physical limitations in PET imaging.

## VII. Conclusion

In this work, we proposed a physically based IR framework for PET image reconstruction, integrating accurate system modeling with automatic differentiation. The framework was evaluated using both phantom and clinical brain datasets. Quantitative and qualitative results demonstrate that our method achieves competitive performance compared to the CASToR reconstruction toolkit and the Siemens Biograph mCT platform, highlighting the potential of the IR framework to enhance quantitative accuracy and anatomical delineation in both research and clinical PET imaging applications.

By combining Monte Carlo and analytical modeling within a differentiable pipeline, our framework supports the incorporation of complex physical processes during reconstruction—without requiring manual gradient derivation. This extensibility makes the proposed framework a promising foundation for advanced reconstruction algorithms that are both interpretable and adaptable.

## Acknowledgment

The authors gratefully acknowledge the assistance of William Calimag at Weill Cornell Medicine for the operation of the Biograph mCT.